\documentclass[journal = jpccck]{achemso}
\usepackage{epsf}
\usepackage{ifthen}
\usepackage[usenames]{color}
\usepackage{amssymb,amsmath,amsbsy}
\usepackage{amsfonts}
\usepackage{amsmath}
\usepackage{graphics, graphicx}
\usepackage{achemso}
\usepackage{bm}

\title{Real-Time Description of the Electronic Dynamics for a Molecule close to a Plasmonic Nanoparticle}

\author{Silvio Pipolo}
\email{silvio.pipolo@impmc.upmc.fr}
\affiliation{Institut de Min\'{e}ralogie, de Physique des Mat\'{e}riaux et de Cosmochimie, Universit\'{e} Pierre et 
Marie Curie - Sorbonne Universit\'{e}s, Paris, France}
\author{Stefano Corni}
\email{stefano.corni@nano.cnr.it}
\affiliation{CNR Istituto Nanoscienze, Modena, Italy}
\SectionNumbersOn

\begin{document}
\renewcommand{\v}[1]{\ensuremath{\mathbf{#1}}} 
\newcommand{\gv}[1]{\ensuremath{\mbox{\boldmath$ #1 $}}} 
\renewcommand{\t}[1]{\ensuremath{\text{#1}}} 
\begin{tocentry}
\includegraphics[width=3.25in]{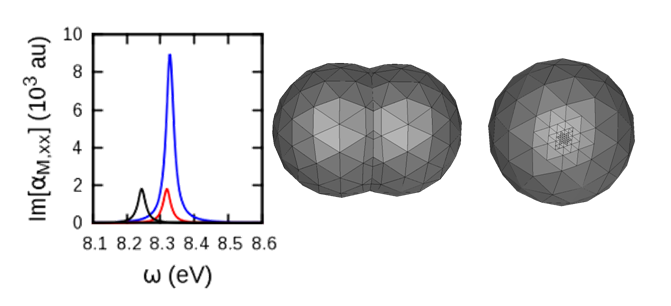}
\end{tocentry}
\begin{abstract}
The optical properties of molecules close to plasmonic nanostructures greatly differ from their isolated molecule counterparts. To theoretically investigate such systems in a Quantum Chemistry perspective, one has to take into account that the plasmonic nanostructure (e.g., a metal nanoparticle - NP) is often too large to be treated atomistically. Therefore, a multiscale description, where the molecule is treated by an ab initio approach and the metal NP by a lower level description, is needed. Here we present an extension of one such multiscale model [Corni, S.; Tomasi, J. {\it J. Chem. Phys.} {\bf 2001}, {\it 114}, 3739] originally inspired by the Polarizable Continuum Model, to a real-time description of the electronic dynamics of the molecule and of the  NP. In particular, we adopt a Time-Dependent Configuration Interaction (TD CI) approach for the molecule, the metal  NP is described as a continuous dielectric of complex shape characterized by a Drude-Lorentz dielectric function and the molecule- NP electromagnetic coupling is treated by an equation-of-motion (EOM) extension of the quasi-static Boundary Element Method (BEM). The  model includes the effects of both the mutual molecule- NP time-dependent polarization and the modification of the probing electromagnetic field due to the plasmonic resonances of the  NP. Finally, such an approach is applied to the investigation of the light absorption of a model chromophore, LiCN, in the presence of a metal   NP of complex shape. 
\end{abstract}

\section{Introduction}\label{Sec:Intro}

The response of a molecule to an applied electromagnetic field is strongly modified by the nearby presence of metal  nanoparticles (NPs) or nanostructures. The resulting effects define the field of {\it molecular plasmonics}\cite{VanDuyne2004}
and include many striking phenomena such as Surface Enhanced Raman Scattering (SERS), Surface Enhanced Infrared Absorption (SEIRA) and Metal Enhanced Fluorescence. Several theoretical models have been proposed through the years to understand such phenomena, some of which exploit a quantum-chemical description of the molecule.\cite{jensen08,morton2011,Chen2012} For these models, the nearby nanostructure is typically too large to be routinely treated by Quantum Chemistry (QM) methods as well. Presently, metal NPs of up to around 1000 atoms, i.e., a few nm in diameter, can be described {{by QM methods}},\cite{Malola2013,Iida2014} {{but they are}} much smaller than many NPs of practical interest. To tackle this limitation, and taking into account the good quality that a classical electromagnetic (EM) description gives of the optical properties of the  NP, in the past different groups devised multiscale methods that keep the quantum mechanical description of the molecule but include the  NP as a classical object, either as a continuous dielectric\cite{Corni2001,chen10c,Mullin2012,Gao2013,Sakko2014} or as a collection of polarizable (and chargeable) atoms.\cite{morton10,morton11a,Payton2012,Rinkevicius2016} 

Most of these models treat the applied electric field, and the corresponding molecule+NP response, in the frequency domain, i.e., assuming a monochromatic sinusoidal EM perturbation acting on the system. While this is in principle a very general approach (any time-dependent EM field can be Fourier-decomposed in a superposition of sinusoidal monochromatic fields), it is not the only one. A fully real-time description of optical phenomena involving molecules is also possible, i.e., a numerical approach that solves the time-dependent Schr\"odinger equation (or the time-dependent Kohn-Sham equation for time dependent density functional theory, TDDFT, approaches) for the molecule in the presence of a time-dependent field. Such real-time description is quite suitable when one wants to simulate the interaction of molecules with short light pulses (that span many frequencies) such as those employed in ultrafast spectroscopies, also in the presence of plasmonic nanostructures.\cite{Vasa2009,Brinks2013,Gruenke2016} It is also a convenient approach for non-linear optical properties simulations.\cite{Castro2006} Finally, a real-time description also offers an alternative way to interpret optical phenomena involving molecules, that may disclose more clearly the physics taking place in the system.\cite{Falke2014} On the down side, real-time description typically requires large computational efforts if fine features in a spectra should be solved, and many of the analysis that are straightforward in the frequency domain (e.g., orbital excitations contributing to a given transition; inspection of the transition density) requires extra steps.

In this work we shall present an evolution of the Polarizable Continuum Model (PCM)\cite{Tomasi2005}-inspired approach that { has} been developed in the past to treat optical properties of molecules close to metal  NP\cite{Corni2001} to the real-time domain. In that model, the molecule was treated initially with Hartree-Fock/time dependent Hartree Fock (HF/TDHF)\cite{Tomasi2005} and then by DFT/TDDFT\cite{corni2002,Andreussi2004} and ZINDO,\cite{Caricato2006a} and was coupled to a classical dielectric description of the metal  NP employing an empirical frequency dependent dielectric constant corrected for quantum-size effects.\cite{Corni2001} The Boundary Element Method (BEM) was used to solve the quasi-static electromagnetic problem needed to write the effective Hamiltonian for the molecule close to the  NP.  Many phenomena were treated with this model, such as SERS,\cite{corni02} SEIRA,\cite{corni06} Metal-Enhanced Fluorescence\cite{Andreussi2004,Caricato2006a,Vukovic2009,Andreussi2013} and Excitation Energy Transfer (EET) close to a NP.\cite{Angioni2013,Andreussi2013} Here the same model will be extended to simulations in real-time. The optical phenomena that will be considered here is the light absorption of the molecule- NP system, with the goal to extend the method to a range of phenomena in the near future.

In the field of quantum-chemical models for molecules close to  NPs, a real time approach has been already used in the past.\cite{chen10c,Gao2013,Sakko2014} The approach that we are proposing here differs from previous ones in two main aspects. The first is the choice of the quantum mechanical method for the molecule. All the cited works exploit real time TDDFT (RT-TDDFT). Here for the molecule we shall instead solve the time dependent Schr\"odinger equation projected on a finite basis of multi-electronic states, obtained by Configuration Interaction (TD CI approach). The second difference is that previous works used the Finite Difference Time Domain (FDTD) method for the  NP polarization, while here we shall exploit BEM (in the quasistatic limit). 

The choice of a wavefunction based method (CI) instead of RT-TDDFT is motivated by the qualitative artifacts that plague the current versions of RT-TDDFT when the molecular state departs substantially from the ground state. For example, the behavior in response to a light pulse that should take most of the population to a given excited state is unsound with respect to the pulse parameter,\cite{Raghunathan2011,Raghunathan2012,Raghunathan2012a}  and  Rabi oscillations are not properly described.\cite{Habenicht2014} The cause of these artifacts is believed to be the adiabatic approximation, i.e., the assumption that the exchange-correlation (xc) potential at a given time depends only on the density at that time rather than to all the previous instants.\cite{Elliott2012,Provorse2016} 
Therefore, although RT-TDDFT will likely become in the future as accurate in the study of electronic dynamics as DFT and TDDFT are now for ground and excited state properties, at present such artifacts make difficult to correctly assign the physical origin of features observed in the electronic dynamics of a molecule close to a metal  NP. TD CI offers the simplest theoretical framework to bypass these problems and has been already applied to the investigation of multi-electronic dynamics for molecules in gas-phase.\cite{krause2005time,Schlegel2007,Krause2008,Kuleff2014} Here, we shall limit ourselves to the simplest CI method, CI Single (CIS). Although CIS is known to have a limited accuracy in terms of excitation energies,\cite{Dreuw2005}  it provides qualitatively correct behaviors also for charge transfer states, and can be systematically improved (e.g., the CIS(D) correction has been already used in a TD context).\cite{krause2005time,Schlegel2007}

The classical electromagnetic description of the  NP via BEM is an alternative approach to FDTD. BEM and FDTD are both widespread methods to study plasmonic nanosystems.\cite{DellaSala2013} Here, BEM has some computational advantages when coupled to a molecule described with a localized basis set, that will be discussed in the theoretical section. BEM is most naturally a technique applied in the frequency domain, and needs to be translated to the time domain to be directly coupled to TD CI. We have recently investigated the electronic dynamics of molecules in solution by a novel extension to the time domain of the Polarizable Continuum Model (PCM) implicit solvation approach.\cite{Corni2014,Pipolo2017} To this aim, we have developed equations of motion (EOMs) for the apparent charges that describe the evolution of time-dependent solvent polarization. Such EOMs are able to describe the delayed solvent response coded in the Debye dielectric constant. Formally, the electrostatic problem involved in PCM and in a quasistatic BEM description of a molecule close to a continuous metal  NP is the same.\cite{Tomasi2002} As such, the EOM method developed in ref. \citenum{Corni2014} can be extended to NPs as well, as anticipated in that work. However, the Debye dielectric constant used there is not appropriate for metal  NPs. Therefore, here we shall develop EOM suitable for the more complex Drude-Lorentz dielectric function,\cite{jackson99} and adopt a numerical approach consistent with such EOMs. The real-time TD CI coupled with EOM BEM evolution is implemented in the stand-alone home-made software WaveT that is interfaced with a local version of the widespread Quantum Chemistry code GAMESS.\cite{Schmidt1993,Gordon2005} 

The article is organized as follows: in section \ref{sec:theo} we shall present the Theoretical basis of the current approach, including the derivation of the EOM for the  NP apparent charges that originates from the Drude-Lorentz dielectric function. Then, in section \ref{sec:res}, we shall apply this theory to study the absorption spectrum of a molecule close to a complex-shaped metal  NP. Finally, we shall draw the conclusions and discuss the development perspectives of this modeling approach.

\section{Theory} \label{sec:theo}
The basic features of the model we are using for a molecule close to a NP has been already described elsewhere.\cite{Corni2001,corni2002,Andreussi2004} Here we summarize them shortly. The  NP is described as a continuous dielectric with a complex shape and  characterized, in the present work, by a Drude-Lorentz dielectric function $\epsilon(\omega)$:\cite{jackson99}
\begin{eqnarray}
\epsilon(\omega)= 1+\frac{A}{\omega_0^2-\omega^2-i\gamma\omega}
\label{eq:Jackson}
\end{eqnarray}
The metal NP is polarized by both the externally applied, incident time-dependent electric field $\vec{E}_\text{inc}(\vec{r},t)$, and by the time-dependent EM field originated by the time-varying electronic charge density of the molecule. In turn, such two polarizations create time-dependent electric fields (sometimes dubbed reflected field $\vec{E}_\text{ref}(\vec{r},t)$ and image or polarization field $\vec{E}_\text{pol}(\vec{r},t)$, respectively)\cite{metiu84,Corni2013a}  that act on the molecule. In the quasistatic limit (i.e., in the limit of electric field wavelength much larger than the  NP)\cite{novotny12} such electric fields are locally associated to time-dependent electrostatic potentials ($V_\text{ref}(\vec{r},t)$ and $V_\text{pol}(\vec{r},t)$) that satisfy proper quasi-static Poisson equations and the electrostatic boundary conditions at the  NP surface.\cite{Corni2001,pipolo2014CEF} Following standard Integral Equation Theory and Boundary Elements Method, (IEF-BEM) each of these electrostatic potentials (reflected and image) can be written as originated by a proper set of point charges ($\mathbf{q}_\text{ref}(t)$ and $\mathbf{q}_\text{pol}(t)$) located in representative points on the  NP surface.\cite{Corni2001}    

The molecule is described by an effective Hamiltonian $\hat{H}(t)$ that includes the electromagnetic interaction with the incident field $\vec{E}_\text{inc}(\vec{r},t)$, the reflected field $\vec{E}_\text{ref}(\vec{r},t)$ and the polarization field $\vec{E}_\text{pol}(\vec{r},t)$. The first is included in the usual dipole approximation, the second and the third interactions are instead written exploiting the charges $\mathbf{q}_\text{ref}(t)$ and $\mathbf{q}_\text{pol}(t)$:
\begin{equation}
 \hat{H}=\hat{H}^0-\hat{\vec{\mu}}\cdot\vec{E}_\text{inc}(\vec{r}_\text{M},t)+({{\v q}}_\text{ref}(t)+{{\v q}}_\text{pol}(t))\cdot\hat{\v V}
\label{eq:H(t)}
\end{equation}
Here, $\hat{H}^0$ is the Hamiltonian of the isolated molecule, $\hat{\vec{\mu}}$ is the dipole operator, $\vec{E}_\text{inc}(\vec{r}_\text{M},t)$ is evaluated in the molecular center of charge $\vec{r}_\text{M}$ and $\hat{\v{V}}$ is a vector operator representing the electrostatic potential of the solute at the representative points on the  NP surface where the apparent charges  $\mathbf{q}_\text{ref}(t)$ and $\mathbf{q}_\text{pol}(t)$ are also located. Notably, ${{\v q}}_\text{pol}(t)$  depend on the electrostatic potential originated by the molecule $V(t')$ at all the previous instants $t'<t$, giving rise to a non-linear and  non-local-in-time evolution problem.\cite{Corni2014}

In the next subsection, we shall focus on the calculations of $\mathbf{q}_\text{ref}(t)$ and $\mathbf{q}_\text{pol}(t)$ for a  NP described by the Drude-Lorentz dielectric constant.

\subsection{Equations of motion for the apparent surface charges for a Drude-Lorentz NP}

Following previous works,\cite{Corni2001,Corni2014} in the frequency domain $\mathbf{q}_\text{pol}(\omega)$ and $\mathbf{q}_\text{ref}(\omega)$ are calculated from the electrostatic potential produced by the molecular charge density on the NP discretized surface (${\v V}(\omega)$) and the potential associated to the incident field $V_\text{inc}(\vec{r})=-\vec{r}\cdot\vec{E}_\text{inc}(\vec{r}_M,\omega)$, respectively, as follow:
\begin{eqnarray}
\nonumber {\v{q}}_\text{pol}(\omega)&=&\v{Q}(\omega)~{\v V}(\omega) \\
\label{eq:qw=QwVw}
{\v{q}}_\text{ref}(\omega)&=&\v{Q}(\omega)~{\v V}_\text{inc}(\omega)
\end{eqnarray}
where $\v{Q}(\omega)$ is the response matrix in the frequency domain:
\begin{eqnarray}
\v{Q}=-\v{S}^{-1}\left( 2\pi \frac{\epsilon(\omega)+1}{\epsilon(\omega)-1}\v{I}+\v{DA} \right)^{-1}\left(2\pi \v{I} +\v{DA} \right)
\label{eq:Qw}
\end{eqnarray}
$\v{A}$ is a diagonal matrix with elements equal to the tessera areas, $\v{S}$ and $\v{D}$ are the representative matrices of the Calderon's projectors\cite{Tomasi2005}:
\begin{eqnarray}
D_{ij}=\frac{(\vec{s}_i-\vec{s}_j)\cdot\vec{n}_j}{|\vec{s}_i-\vec{s}_j|^3} \qquad S_{ij}=\frac{1}{|\vec{s}_i-\vec{s}_j|}
\end{eqnarray}
$\vec{s}_i$ are the representative points on the nanoparticle surface and $\vec{n}_j$ are unit vectors normal to the j-th tessera and pointing outwards.
The generalization to the time domain of Eq. (\ref{eq:qw=QwVw}) is straightforward (for the sake of brevity, we focus here on $\mathbf{q}_\text{pol}$ and ${\v V}$ only, the equations are identical for the pair $\mathbf{q}_\text{ref}$ and ${\v V}_\text{inc}$):
\begin{eqnarray}
\v{q}_\text{pol}(t)=\int^{\infty}_{-\infty}\v{Q}(t-t')~{\v V}(t')~ dt'
\end{eqnarray}\label{eq:qt=Qtvt}
with charges and potentials defined in the time domain and $\v{Q}(t-t')$ being the kernel (or memory) matrix that determines the non-equilibrium response of the  NP in the time domain, obtained by Fourier-transforming $\v{Q}(\omega)$.
Following the route presented in ref. \citenum{Corni2014}, we rewrite the $\v{Q}(t-t')$ matrix in terms of a diagonal representation:
\begin{eqnarray}
\mathbf{q}_\text{pol}(t)&=&- \int_{-\infty}^{\infty} dt'~ \mathbf{S}^{-1/2}\mathbf{T} \mathbf{K}(t-t') \mathbf{T}^{\dagger}\mathbf{S}^{-1/2}\mathbf{V}(t') \label{eq:IEF_q}\\
K_{ii}(t-t')&=&\int_{-\infty}^{\infty} \frac{d \omega}{2\pi} e^{-i\omega(t-t')} K_{ii}(\omega)~~~~K_{ii}(\omega)=\frac{2\pi+\Lambda_{ii}}{2\pi\frac{\epsilon(\omega)+1}{\epsilon(\omega)-1}+\Lambda_{ii}}
\label{eq:diag_ft}\\
K_{ij}(t-t')&=&0~~i\neq j
\end{eqnarray}
where the matrices $\mathbf{T}$ and 
$\boldsymbol{\Lambda}$ are build by the eigenvectors and the eigenvalues, respectively, of the symmetric matrix $1/2\left( \mathbf{S}^{-1/2}\mathbf{DA}\mathbf{S}^{1/2}+\mathbf{S}^{1/2}\mathbf{AD}^{\dagger}\mathbf{S}^{-1/2}\right) \approx \mathbf{S}^{-1/2}\mathbf{DA}\mathbf{S}^{1/2}$ { (the identity is exact for the corresponding integral operators)}. 

Taking into account the specific form of the Drude-Lorentz dielectric function Eq.(\ref{eq:Jackson}), $K_{ii}(\omega)$ in Eq.(\ref{eq:diag_ft}) can be written as:
\begin{eqnarray}
 \nonumber K_{ii}(\omega)&=&\frac{K_{f,ii}}{\omega_0^2-\omega^2-i\gamma\omega+K_{f,ii}}\\
 K_{f,ii}&=&\frac{(2\pi+\Lambda_{ii})A}{4\pi}
\label{eq:diag_Kw}
\end{eqnarray}
We compute then the Fourier transform 
\begin{eqnarray}
K_{ii}(t-t')&=&\int_{-\infty}^{\infty} \frac{d \omega}{2\pi}~  \frac{K_{f,ii}}{\omega_0^2-\omega^2-i\gamma\omega+K_{f,ii}}e^{-i\omega(t-t')}  
\end{eqnarray}
to obtain the following form of the diagonal kernel function:
\begin{eqnarray}
K_{ii}(t-t')&=&\frac{K_{f,ii}}{\bar{\omega}_{0,ii}}e^{-\frac{(t-t')}{\tau}}\sin{[\bar{\omega}_{0,ii}(t-t')]}\Theta(t-t')
\label{eq:diag_Kt}  
\end{eqnarray}
with
\begin{eqnarray}
\bar{\omega}_{0,ii}&=&\sqrt{-\frac{\gamma^2}{4}+K_{f,ii}+\omega_0^2}  \\
\tau&=&\frac{2}{\gamma}  
\end{eqnarray}
{{$\Theta(t-t')$ is the Heaviside step function.}}
The equations for the apparent charges are then obtained by substituting in Eq.(\ref{eq:IEF_q}) the expression of the diagonal kernel 
function $K_{ii}(t-t')$ given in Eq.(\ref{eq:diag_Kt}).

It is now possible to write an EOM for the charges $\mathbf{q}_\text{pol}(t)$. The procedure consists in taking the first and the second derivatives w.r.t. time of Eq. (\ref{eq:IEF_q}), and use them to eliminate the convolution integral from the definition of $\mathbf{q}_\text{pol}(t)$, similarly to what done in ref. \citenum{Corni2014}. This leads to the EOM:

\begin{eqnarray}
\label{eq:eom}\mathbf{\ddot{q}}_\text{pol}(t)&=&-\gamma\mathbf{\dot{q}}_\text{pol}(t) - \mathbf{Q}_{\omega} \mathbf{q}_\text{pol}(t)  + \mathbf{Q}_f\mathbf{V}(t)\\
&& \mathbf{Q}_{\omega}= \mathbf{S}^{-1/2}\mathbf{T} \mathbf{K}_{\omega} \mathbf{T}^{\dagger}\mathbf{S}^{1/2}\\
&& \mathbf{Q}_{f}= -\mathbf{S}^{-1/2}\mathbf{T} \mathbf{K}_{f} \mathbf{T}^{\dagger}\mathbf{S}^{-1/2}
\end{eqnarray}
and analogously for the reflected field charges:
\begin{eqnarray}
\label{eq:eom_f}\mathbf{\ddot{q}}_\text{ref}(t)&=&-\gamma\mathbf{\dot{q}}_\text{ref}(t) - \mathbf{Q}_{\omega} \mathbf{q}_\text{ ref}(t)  + \mathbf{Q}_f\mathbf{V}_\text{inc}(t)
\end{eqnarray}

with $\mathbf{K}_{f}$ being the diagonal matrix defined before in Eq.(\ref{eq:diag_Kw}) and $\mathbf{K}_{\omega}$ still diagonal and with elements defined by:
\begin{eqnarray}
\label{eq:freq} K_{\omega,ii}&=&\bar{\omega}_{0,ii}^2 + \frac{\gamma^2}{4}=\frac{(2\pi+\Lambda_{ii})A}{4\pi}+\omega_0^2=K_{f,ii}+\omega_0^2
\end{eqnarray}

The form of Eqs.(\ref{eq:eom}) and (\ref{eq:eom_f}) is clearly that of (coupled) forced damped harmonic oscillators ($\ddot{x}=-\gamma\dot{x}-\omega_{res}^2 x+F$). The squared frequencies of the oscillators are those given in Eq.(\ref{eq:freq}), and depend on two quantities: one is the Lorentz resonance frequency $\omega_0$ characteristic of the material making up the  NP, the other ($K_{f,ii}$) is a purely geometric term that is related to the shape of the  NP and the various excitations it can sustain. We have already discussed the possible values of $\Lambda_{ii}$ for a spherical  NP in ref. \citenum{Corni2014}. Based on those results, we find that for a spherical metal  NP described by a Drude dielectric constant (i.e., $\omega_0=0$ and $A=\Omega_p^2$ in Eq.(\ref{eq:Jackson}), where $\Omega_p$ is the plasma frequency), the squared frequencies of the plasmonic modes (i.e., $K_{\omega,ii}$) are:

\begin{eqnarray}
K_{\omega,lm}^{\text Drude}&=&\frac{\Omega_p^2}{1+(l+1)/l},~~l>0,~~m=-l,...,l
\end{eqnarray}
i.e., the well-known frequencies of the multipolar plasmons of a spherical particle. 

From a practical point of view, numerically integrating the EOM Eq.(\ref{eq:eom}) requires some care.\cite{swope1982computer} Here we exploit the velocity-Verlet scheme proposed in ref. \cite{Vandeneijnden2006}, second-order accurate in the numerical integration step $dt$.

Extension to dielectric constant made of sums of Drude-Lorentz terms is straightforward, as they {create a system of coupled oscillators}.   

\subsection{Coupling to TD CI}
The theory of TD CI in presence of a solvent with a delayed response has been recently developed.\cite{Pipolo2017} The equivalent TD CI theory for a molecule coupled to a metal  NP is very similar, and will be only briefly summarized here.

The state vector $|\Psi(t)\rangle $ of the molecule satisfies the time-dependent non-linear Schr\"{o}dinger equation
\begin{equation}
 i\frac{\partial}{\partial t}|\Psi(t)\rangle =\hat{H}|\Psi(t)\rangle 
\label{TDNLSE}
\end{equation}
Approximated solutions can be obtained by writing $|\Psi(t)\rangle$ as a linear combination with time-dependent coefficients $C_I(t)$ of a reference ground state $|\Phi_0\rangle $ and a finite set of excited states $|\Phi_I\rangle $ 
\begin{equation}
 |\Psi\rangle = \sum_{I} C_I(t)|\Phi_I\rangle 
 \label{eq:base}
\end{equation}

By using Eq. (\ref{eq:base}), the time-dependent Schr\"{o}dinger equation, Eq. (\ref{TDNLSE}), becomes: 
 \begin{equation}
  i \frac{d {\v C}}{dt}={\v H} {\v C}
  \label{eq:schro}
 \end{equation}
where the Hamiltonian matrix ${\v H}$ has elements:
\begin{equation}
 H_{IJ}(t)=\langle \Phi_I|\hat{H}^0-\hat{\vec{\mu}}\cdot\vec{E}_\text{inc}(\vec{r}_\text{M},t)+({{\v q}}_\text{ref}(t)+{{\v q}}_\text{pol}(t))\cdot\hat{\v V}|\Phi_J\rangle   
\end{equation}

We choose a reference state given by the Hartee-Fock determinant $|\Phi_0\rangle =|\text{HF}\rangle $ for the molecule equilibrated with the  NP described as a perfect (and charge neutral) conductor, and we use as the excited states $|\Phi_I\rangle $ (with $I> 0$) those obtained from a configuration interaction expansion limited to single excitations (CIS). Such CIS excited states $|\Phi_I\rangle $ and the corresponding energies are obtained here by solving, in the subspace of the single excited determinants $|\psi_i^a\rangle =\hat{a}^\dag\hat{i}|\text{HF}\rangle$ {{($\hat{a}^\dag$ and $\hat{i}$ are electron creation and annihilation operators respectively)}}, the time-independent Schr\"{o}dinger equation for  the molecular solute in the presence of the fixed Hartee-Fock polarization charges:
\begin{equation}
\left [ \hat{H}^0+{  {\v q}_\text{pol}}(|\text{HF}\rangle)\cdot\hat{\v V} \right ]|\Phi_I\rangle =E_I|\Phi_I\rangle 
\end{equation}

With this choice, the elements of the Hamiltonian matrix in Eq. (\ref{eq:H(t)}) become:
\begin{eqnarray}
H_{IJ}(t)&=&E_I\delta_{IJ}-\vec{\mu}_{IJ}\cdot\vec{E}_\text{inc}(\vec{r}_\text{M},t)+({{\v q}}_\text{ref}(t)+\Delta{{\v q}}_\text{pol}(t))\cdot{\v V}_{IJ}
\label{eq:H_IJ}
\end{eqnarray}
where $\Delta{  {\v q}}_\text{pol}(t)$, $\vec{\mu}_{IJ}$ and ${\v V}_{IJ}$ are defined by
\begin{eqnarray}
{\Delta {\v q}_\text{pol}}(t)&=& {\v q}_\text{pol}(t)-  {\v q}_\text{pol}(|\text{HF}\rangle)
\label{eq:int}\\
\vec{\mu}_{IJ}&=&\langle \Phi_I |\hat{\vec{\mu}}|\Phi_J\rangle\\
{\v V}_{IJ}&=&\langle \Phi_I |\hat{{\v V}}|\Phi_J\rangle
\end{eqnarray} 

${\v q}_\text{pol}(t)$ in Eq.(\ref{eq:int}) and ${\v q}_\text{ref}(t)$ in Eq.(\ref{eq:H_IJ}) are obtained by the EOM approach described in the previous section. Numerically, Eq.(\ref{eq:schro}) is solved by a simple Euler method at the first order,\cite{Schlegel2007} using the same $dt$ as for the apparent charge EOMs.

We close this section by remarking that within the present BEM approach, the electrostatic potentials ${\v V}_{IJ}$ in Eq.(\ref{eq:H_IJ}) have to be calculated only for points on the NP surface. In a FDTD approach, they would be needed on a volume including the NP, making the calculation rather cumbersome.

\subsection{Computational Details}
The geometry of the LiCN molecule used in the calculations is the one of ref.\citenum{krause2005time} (LiC and CN bonds lengths of $1.949$ \AA{} and $1.147$ \AA{} respectively).  
We have used a Drude dielectric function (i.e., $\omega_0=0$) with parameters for silver:\cite{Corni2001}  plasma frequency $\Omega_p = 0.332$au $={ 9.03}$eV ($A=\Omega_p^2$) and $\gamma=1.515 \cdot 10^{-3}$au $=4.123 \cdot 10^{-2}$eV. No interband transition is included in the dielectric constant, therefore the NP absorption spectra are not directly comparable to silver NP experimental ones.  
The calculation of the HF and CIS states in the presence of the  NP, their energies (at frozen  NP dielectric response), the expectation values and transition integrals of the dipole moment and the electrostatic potential operators, are performed using the 6-31G(d) basis set \cite{hariharan1973influence} exploiting a locally modified version of the GAMESS code.\cite{Schmidt1993,Gordon2005}  Only the first 15 excited states are kept in the Hamiltonian.
The coupled propagation of the wavefunction of the molecule, the polarization and reflected field charges is performed using Wave-T, an home-made code. The time step of the propagation simulations is $dt=4.838$as ($0.2$au). For the calculations with the spherical NP, 240 tesseras were used; in those with the elongated NP, the number of tesseras was 472 { (tests with 676 and 850 tesseras were also done). We could perform 10.9 (7.4) ps/day of simulation on a Intel Xeon E5-2650 core and 472 (676) tesseras (no parallelization of the code was attempted)}.

\section{Results and Discussion} \label{sec:res}
\subsection{Tests for the polarization of a spherical  NP}
To test our approach we present the results of the simulations for the induced dipole of the NP in two cases where an analytical result is available, i.e. (i) a spherical  NP subject to a monochromatic time dependent electric field and (ii) a spherical  NP interacting with a static dipole placed at a large distance.  

\begin{figure}
\centering  \includegraphics[width=0.6\textwidth]{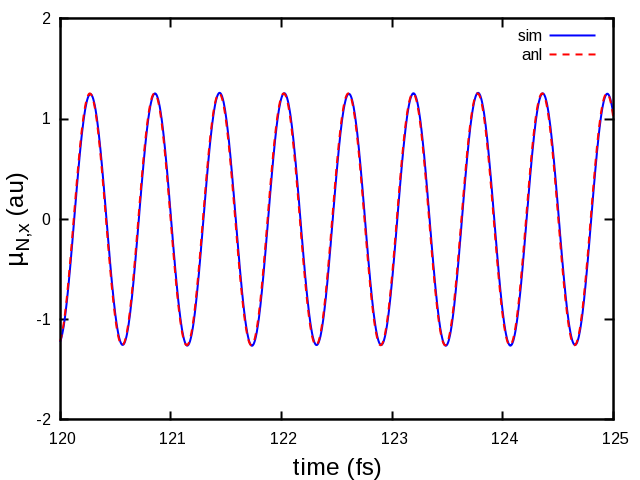} 
\caption{Simulation profiles of the $x$-component of $\vec{\mu}_\text{N}$ for a spherical  NP with a radius of $2.5$nm. The labels ``sim'' and ``anl'' refer to the results of the simulation and to the analytical profile, Eq. (\ref{eq:mu_np_local_anl}), respectively. The frequency of the incident field is $\omega_\text{inc}=0.26$au.}
\label{fig:localF} 
\end{figure}

(i) The analytical expression for the induced dipole moment of the NP ($ \vec{\mu}_\text{N}$) generated by a monochromatic time dependent electric field is the following\cite{bottcher1973theory}:
\begin{eqnarray}\label{eq:mu_np_local_anl}
\vec{\mu}^\text{anl}_\text{N}(t)=\frac{\epsilon(\omega_\text{inc})-1}{\epsilon(\omega_\text{inc})+2}a^3 E_\text{inc}~\vec{\varepsilon}~ e^{-i\omega_\text{inc} t} ~ + ~ c.c.
\end{eqnarray}  
where $a$ is the radius of the  NP, $E_\text{inc}$ is the complex amplitude of the incident field, $\vec{\varepsilon}$ is a unit vector defining its direction, and $\epsilon(\omega_\text{inc})$ is the dielectric function evaluated at the frequency $\omega_\text{inc}$. 
In Figure \ref{fig:localF} we compare the profile of $\vec{\mu}_\text{N}(t)$ obtained from the EOM BEM simulation
\begin{eqnarray}\label{eq:mu_np_local_sim}
  \vec{\mu}^\text{sim}_\text{N}(t)= \sum_i q_{\text{ref},i} (t) ~ \vec{s}_i
\end{eqnarray}
with the analytical one, Eq. (\ref{eq:mu_np_local_anl}), in the case of a sinusoidal external field with maximum amplitude of $10^{-5}$au and a $2.5$nm-radius  NP. $\vec{s}_i$ are the positions of the tessera representative points.
During the simulation the incident electric field is slowly turned on by modulating the sinusoidal form with a linear function between $t=0$fs and $t= 120$fs. The frequency of the external field is $\omega_\text{inc}=0.26$au{=7.1eV}. 
The simulated profile and the analytical reference perfectly agree.

(ii) When a spherical  NP is polarized by a static dipole at large distance, $\vec{\mu}_\text{N}$ may be evaluated by taking the $t\rightarrow 0$ limit in Eq. (\ref{eq:mu_np_local_anl}) and considering, as an incident field, the electric field generated by the non polarizable dipole  moment at the center of the  NP. In a simulation we model such a limit case by placing the molecule far away from the  NP and by freezing its charge density as it was in vacuum. The value of $|\vec{\mu}_\text{N}|$ computed for a system's configuration where the center of the  NP is placed at a distance of $25$nm from the center of charge of the molecule is $9.4184 \cdot 10^{-3}$D. The direction of the molecular dipole is perpendicular to the line joining the center of the  molecule and the NP (the  NP is the same used in the previous test). Taking into account that the magnitude of the dipole moment of LiCN in vacuum is $9.425$D, the difference between the calculated and the analytical value is less than $0.1\%$.

These simple tests back up the numerical accuracy of the model. In the next section we shall apply it to a physical problem.

\subsection{LiCN physisorbed on an elongated  NP: absorption spectra}\label{sec:abs}
\begin{figure}
\centering  \includegraphics[width=0.6\textwidth]{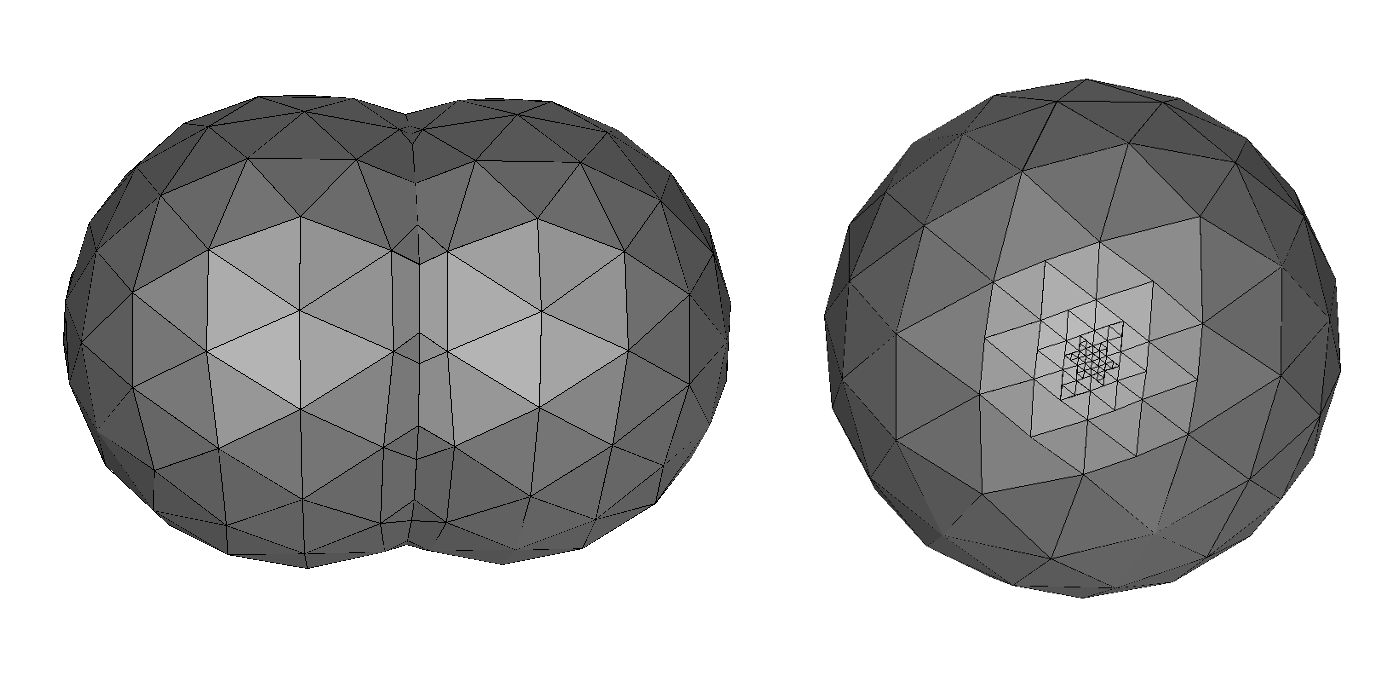} 
\caption{Side (left) and front (right) view of the  NP used in section \ref{sec:abs}. The long axis of the NP is directed along $x$. The  NP surface is divided in 472 tesseras, refined in the region closer to the molecule.
} 
\label{fig:np} 
\end{figure}

\begin{figure}
\centering  \includegraphics[width=0.6\textwidth]{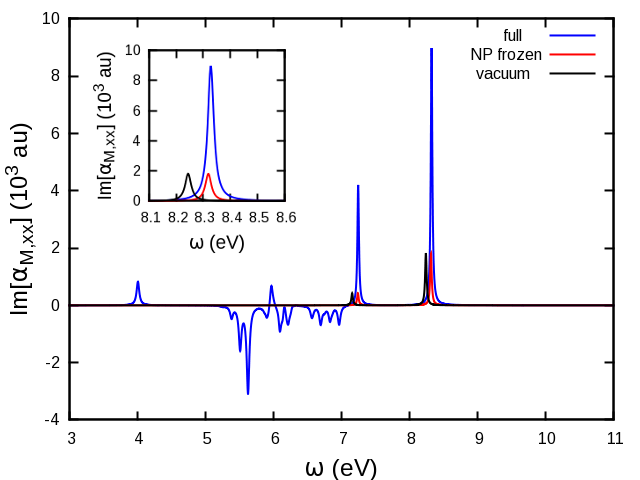} 
\caption{The imaginary part of $\alpha_{\text{M},xx}$ is plotted as a function of frequency for (i) a simulation of the molecule in vacuum (vacuum), (ii) a simulation with the  NP polarization frozen in its equilibrium value when no perturbation is acting on the system (NP frozen), and (iii) a simulation with the full propagation of molecule and  NP degrees of freedom (full - see section \ref{sec:theo}). Inset: a zoom in the $8.1-8.6$eV region.} 
\label{fig:mol_abs} 
\end{figure}

\begin{figure}
\centering  \includegraphics[width=0.6\textwidth]{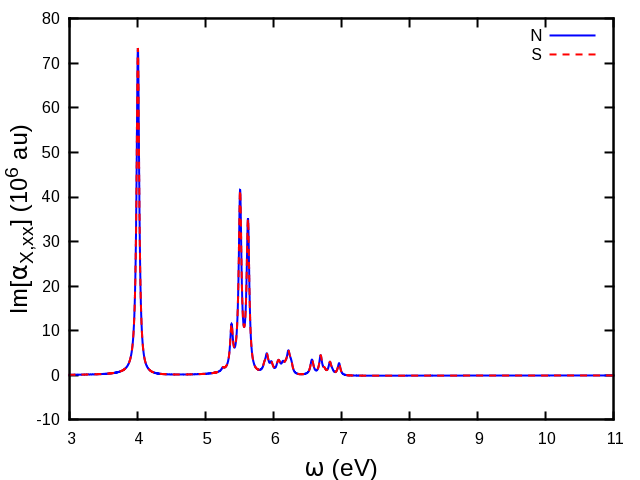} 
\caption{The imaginary parts of $\alpha_{\text{S},xx}$ (S) and $\alpha_{\text{N},xx}$ (N) are plotted as a function of frequency for a simulation where the full propagation of molecule and  NP degrees of freedom is performed (see section \ref{sec:theo}).} 
\label{fig:molnp_abs} 
\end{figure}

In this section we discuss the results obtained from {{EOM BEM}} TD CIS simulations of a LiCN-NP system where the metal  NP is build as the union of two intersecting sphere, each with radius 5nm and having centers displaced of 4nm along the $x$ direction. The LiCN molecule is also placed along $x$, with the N atom closest to the NP and at 4 {\AA} from its surface. The surface of the  NP is divided in tesseras that are smaller closer to the molecule (see Fig.\ref{fig:np}).
We compute the frequency-dependent linear polarizability tensors ($\bar{\alpha}_{\t X}$) from the Fourier transforms of the induced dipoles signals for the molecule ($\t X= \t M$), the  NP ($\t X= \t N$) and the entire system ($\t X= \t S$). The elements of $\bar{\alpha}_{\t X}$ may be written as follows\cite{Pipolo2014}
\begin{eqnarray}\label{eq:alpha}
\alpha_{\text{X},jk}(\omega)=\frac{1}{2\pi E_{\text{inc},k}(\omega)} \int_0^{\infty} \Delta \mu_{\text{X},j} (t) ~ e^{i(\omega+\frac{i}{\tau_\text{M}})t}~dt
\end{eqnarray}  
here $j$ and $k$ refer to two of the three directions in Cartesian space, $E_{\text{inc},k}(\omega)$ is the component of the Fourier transform of the perturbing field along $k$, $\Delta\mu_{\text{X},j}(t)=\mu_{\text{X},j}(t)-\mu_{\text{X},j}(0)$ is the component of the induced dipole profile along $j$ and $\tau_\text{M}$ is a damping constant that allows to model a finite lifetime for the excited states ($\tau_\text{M}=2000~$au$={ 48.38}$ fs).\cite{Pipolo2014} 
A narrow Gaussian pulse, mimicking a $\delta$-pulse, directed along the $x$ axis is used to perturb the system within the linear regime. The perturbing field has a width of $48.38$as and a maximum intensity of $10^{-6}$au. The imaginary part of $\alpha_{\text{X},xx}(\omega)$ is proportional to the absorption cross section.\cite{novotny12} 
In Figure \ref{fig:mol_abs} we compare such imaginary parts for the molecular polarizabilities in vacuum, and close to the  NP. Two simulations have been performed to address the effect of the  NP polarization on the molecular properties, one with the  NP polarization frozen at its equilibrium value before the action of the perturbation ($\Delta \v{q}_\text{pol}(t)=\v{q}_\text{ref}(t)=0$; ``NP frozen'' in Figure \ref{fig:mol_abs}), and one with the full propagation of both the NP and molecule electronic dynamics as detailed in section \ref{sec:theo} (``full'' in Figure \ref{fig:mol_abs}).
The effect of the equilibrium polarization of the  NP manifests as a blue shift on the  NPs excitation energies in gas phase. This is reasonable, since the dipole moments in the excited states corresponding to these excitations are lower than in the ground state, and therefore they are less stabilized by the NP presence.
In addition to this, accounting for the full time-dependent  NP polarization results in a further, but smaller, blue-shift of the molecular excitations and a strong enhancement of the peak intensities (see the inset in Figure \ref{fig:mol_abs}). 
The new peaks at a frequency lower than $7$eV are reminiscent of the NP absorption. They can be negative, as the physical requirement of a positive absorption (starting from the ground state) refers to the overall (molecule+NP) system. 
In Figure \ref{fig:molnp_abs} we report the imaginary parts of the  NP and system polarizabilities for the simulation with the full propagation of the system degrees of freedom. Because of the difference in size between the  NP and the molecule the responses have different orders of magnitude and the effect of the molecule on the NP response is not detectable.

\begin{figure}
\centering  \includegraphics[width=0.6\textwidth]{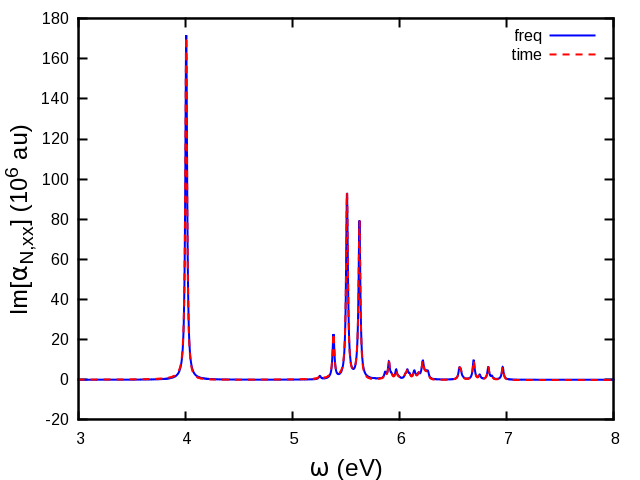} 
\caption{Imaginary part of $\alpha_{\text{N},xx}$ as a function of frequency calculated using Eq. (\ref{eq:alpha}) (time) and directly in the frequency domain (freq).} 
\label{fig:np_abs} 
\end{figure}

Finally in Figure \ref{fig:np_abs} we compare the profiles of the imaginary part of $\alpha_{\text{N},xx}$ for the nanoparticle in vacuum calculated (i) using Eq. (\ref{eq:alpha}) and (ii) directly in the frequency domain. A good agreement is found between the two methods. { We close this section by remarking that the results, in general, will depend on the accuracy of the nanoparticle tessellation. This is discussed in the Supporting Information, where it is shown that increasing the number of tesseras from 472 to 676 can shift the nanoparticle peaks by 0.1-0.2 eV and that peak intensities can also change. In practical applications where specific experiments are targeted, it is therefore necessary to check the convergence of the numerical results with the tesselation.} 

\section{Conclusions and Perspectives}
In this work, we have presented a real-time methodology to study the electronic dynamics of a molecule close to a metal  NP, based on a polyelectronic (TD CI) description of the molecule and a EOM BEM treatment of the molecule-metal  NP electromagnetic coupling (in the quasi-static limit).  We have then applied this approach to study the light absorption by a Drude metal  NP plus molecule system. The present work represents an initial step to provide a comprehensive modeling of the real-time dynamics of molecules close to metal  NPs, that will require developments in various directions. For what regards the molecule, the quality of the wavefunction approach used here can be improved by obtaining the necessary quantities (excitation energies, dipole and electrostatic potential matrix elements) via methods more accurate than CIS, that can be coded in richly featured and flexible Quantum Chemistry software such as GAMESS.  The coupling with RT-TDDFT, notwithstanding its current drawbacks related to the adiabatic approximation, is also important to benefit from the advantageous accuracy vs computational cost balance of such method. Moreover, here the molecular nuclei have been  frozen, that is suitable if only the electronic dynamics is of interest. The study of several other phenomena ranging from vibrational spectroscopy to photochemistry, requires such assumption to be relaxed. 

On the side of the  NP description, there are some extensions that are relatively straightforward, such as to consider more complex dielectric functions that involve a sum of Drude-Lorentz  terms. With this kind of $\epsilon(\omega)$, one can reproduce quite faithfully empirical permittivities { that also account for interband transitions (disregarded here)}, as done in FDTD approaches. Corrections for the limited mean free-path of the electrons in the confining  NP can be included. Non-local aspects of the metal-NP response { (that are included for example in proper non local dielectric functions\cite{mcmahon09})} may be { also included at different levels} in a IEF BEM approach.\cite{corni2002,Christensen2016} In perspective, a supermolecular description of the molecule + NP system may also be envisaged, where part of the  NP is treated quantum-mechanically together with the molecule.\cite{Gao2013} 

Here we have exploited the quasi-static approximation, that becomes less and less accurate for increasing NP size.\cite{Bergamini2013} BEM can be extended to a full electromagnetic description as well,\cite{GarciadeAbajo2002,Hohenester2012} that is still compatible with coupling with a Quantum Chemistry molecule.\cite{pipolo2014CEF} Finally, we have neglected here the possible presence of a solvent/dielectric matrix, that can be included as an additional dielectric (with its own time-dependent response) hosting the molecule in a cavity as done previously in the frequency domain.\cite{Corni2001}  

In conclusion, the development presented here represents a promising starting point to investigate the optical properties of molecules close to metal NPs, that is amenable to various developments currently on-going in our group.

\acknowledgement
We would like to thank Roberto Cammi for very useful discussions and comments.  SC acknowldges funding from the ERC under the grant ERC-CoG-681285 TAME-Plasmons. SC and SP acknowldge the Short-Term Mobility program of the CNR.

\begin{suppinfo}
A discussion on the effect of the NP tessellation.
\end{suppinfo}

\newpage
\providecommand{\latin}[1]{#1}
\providecommand*\mcitethebibliography{\thebibliography}
\csname @ifundefined\endcsname{endmcitethebibliography}
  {\let\endmcitethebibliography\endthebibliography}{}

\end{document}